\documentclass[journal]{IEEEtran}
\IEEEoverridecommandlockouts
\usepackage{cite}
\usepackage{amsmath,amssymb,amsfonts}
\usepackage{algorithmic}
\usepackage{graphicx}   
\usepackage{textcomp}
\usepackage{xcolor}
\usepackage{tikzscale}
\usepackage{booktabs,url}
\usepackage{tabularx}
\usepackage{balance}
\usepackage{multirow}
\usepackage{graphics}
\usepackage{algorithm}
\usepackage[list=true]{subcaption}
\usepackage[font={footnotesize}]{caption}
\def\lf{\left\lfloor} 
\def\rf{\right\rfloor}

\def\BibTeX{{\rm B\kern-.05em{\sc i\kern-.025em b}\kern-.08em
    T\kern-.1667em\lower.7ex\hbox{E}\kern-.125emX}}
\usepackage{acro}
\usepackage{pgfplots}
  \pgfplotsset{compat=newest}
  \usepackage{pgfplotstable}
  \usepgfplotslibrary{groupplots}
  \usetikzlibrary{plotmarks}
  \usetikzlibrary{arrows.meta}
  \usetikzlibrary{calc,positioning}
  \usepgfplotslibrary{patchplots}
  \usepackage{grffile}
  \pgfplotsset{plot coordinates/math parser=false}
  \newlength\figureheight
  \newlength\figurewidth
\DeclareAcronym{5g}{
short=5G,
long= fifth generation,
}
\DeclareAcronym{iq}{
short=I/Q,
long= quadrature,
}
\DeclareAcronym{I}{
short=I,
long= in-phase,
}
\DeclareAcronym{Q}{
short=Q,
long= quadrature,
}
\DeclareAcronym{ls}{
short=LS,
long= least squares,
}
\DeclareAcronym{rvftdnn}{
short=RVFTDNN,
long= real-valued focused time-delay neural network,
}
\DeclareAcronym{lo}{
short=LO,
long= local oscillator,
}
\DeclareAcronym{lpf}{
short=LPF,
long= lowpass filter,
}
\DeclareAcronym{cdf}{
short=CDF,
long= cumulative distribution function ,
}
\DeclareAcronym{fir}{
short=FIR,
long= finite impulse response,
}
\DeclareAcronym{rhs}{
short=RHS,
long= right-hand side,
}
\DeclareAcronym{dsp}{
short=DSP,
long= digital signal processing,
}
\DeclareAcronym{nn}{
short=NN,
long= neural network,
}
\DeclareAcronym{mlp}{
short=MLP,
long=multilayer perceptron
}
\DeclareAcronym{GaN}{
short=GaN,
long=Gallium Nitride,
}
\DeclareAcronym{relu}{
short=ReLU,
long = rectified linear unit, 
}
\DeclareAcronym{mse}{
short=MSE,
long=mean squared error,
}
\DeclareAcronym{rvtdnn}{
short=RVTDNN,
long= real-valued time-delay neural network,
}
\DeclareAcronym{arvtdnn}{
short=ARVTDNN,
long= augmented real-valued time-delay neural network,
}
\DeclareAcronym{arden}{
short=ARDEN,
long= attention residual real-valued time-delay neural network,
}
\DeclareAcronym{r2tdnn}{
short=R2TDNN,
long= residual real-valued time-delay neural network,
}
\DeclareAcronym{flop}{
short=FLOP,
long= floating point operations,
}
\DeclareAcronym{ph}{
short=PH,
long= parallel Hammerstein
}
\DeclareAcronym{ofdm}{
short=OFDM,
long=orthogonal frequency division multiplexing,
}
\DeclareAcronym{par}{
short=PAR,
long=peak-to-average ratio,
}
\DeclareAcronym{papr}{
short=PAPR,
long=peak-to-average power ratio,
}
\DeclareAcronym{rf}{
short=RF,
long=radio frequency,
}
\DeclareAcronym{pa}{
short=PA,
long=power amplifier,
}
\DeclareAcronym{pas}{
short=\acs{pa}s,
long=power amplifiers,
}
\DeclareAcronym{psd}{
short=PSD,
long= power spectral density,
}
\DeclareAcronym{dpd}{
short=DPD,
long=digital predistortion,
}
\DeclareAcronym{cfr}{
short=CFR,
long=crest factor reduction,
}
\DeclareAcronym{cf}{
short=CF,
long=crest-factor}

\DeclareAcronym{evm}{
short=EVM,
long=error vector magnitude,
}
\DeclareAcronym{nmse}{
short=NMSE,
long=normalized mean square error,
}
\DeclareAcronym{acpr}{
short=ACPR,
long=adjacent channel power ratio,
}
\DeclareAcronym{pae}{
short=PAE,
long=power added efficiency,
}
\DeclareAcronym{dla}{
short=DLA,
long=direct learning architecture,
}
\DeclareAcronym{ila}{
short=ILA,
long=indirect learning architecture,
}
\DeclareAcronym{ilc}{
short=ILC,
long=iterative learning control ,
}
\DeclareAcronym{cfr-dpd}{
short=CFR-DPD,
long=CFR combined with DPD,
}
\DeclareAcronym{icf}{
short=ICF,
long=iterative clipping and filtering,
}
\DeclareAcronym{am/am}{
short=AM/AM,
long=amplitude-to-amplitude,
}
\DeclareAcronym{am/pm}{
short=AM/PM,
long=amplitude-to-phase,
}
\DeclareAcronym{siso}{
short=SISO,
long=single-input single-output
}
\DeclareAcronym{mimo}{
short=MIMO,
long=multiple-input multiple-output
}
\DeclareAcronym{mp}{
short=MP,
long=memory polynomial
}
\DeclareAcronym{gmp}{
short=GMP,
long=generalized memory polynomial
}
\DeclareAcronym{adc}{
short=ADC,
long= analog-to-digital converter}
\DeclareAcronym{dac}{
short=DAC,
long= digital-to-analog converter}
\DeclareAcronym{ilc-dpd}{
short=ILC-DPD,
long= adaptive ILC-based DPD
}
\DeclareAcronym{rms}{
short=RMS,
long= root mean squares
}
\DeclareAcronym{vst}{
short=VST,
long= vector signal transceiver
}
\DeclareAcronym{mmwv}{
short=mm-Wave,
long= millimeter-wave
}

\begin{document}

\title{Low Complexity Joint Impairment Mitigation of \\  I/Q Modulator and PA  Using Neural Networks
\thanks{
This work was presented in part at the IEEE Global Communications Conference (GLOBECOM), Taipei, Taiwan~\cite{wu2020residual}.

Y. Wu is with Ericsson Research and Chalmers University of Technology, Gothenburg, Sweden (email: yibo@chalmers.se)

U. Gustavsson is with Ericsson Research, Gothenburg, Sweden (e-mail: ulf.gustavsson@ericsson.com) 

A. Graell i Amat and H. Wymeersch are with Chalmers University of Technology, Gothenburg, Sweden (alexandre.graell@chalmers.se; henkw@chalmers.se)

This work was supported by the Swedish Foundation for Strategic Research (SSF), grant no.~I19-0021. )
}
}
\author{Yibo~Wu,~\IEEEmembership{Student~Member,~IEEE},
        Ulf~Gustavsson,~\IEEEmembership{Senior~Member,~IEEE}, \\
        Alexandre~Graell~i~Amat,~\IEEEmembership{Senior~Member,~IEEE}, and
        Henk~Wymeersch,~\IEEEmembership{Senior~Member,~IEEE}
        }
\maketitle

\begin{abstract}
\Acp{nn} for multiple hardware impairments mitigation of a realistic direct conversion transmitter are impractical due to high computational complexity. We propose two methods to reduce complexity without significant performance penalty. We first propose a novel attention residual learning \ac{nn}, referred to as \acf{arden}, where trainable neuron-wise shortcut connections between the input and output layers allow to keep the attention always active. Furthermore, we implement a \ac{nn} pruning algorithm that gradually removes connections corresponding to minimal weight magnitudes in each layer. Simulation and experimental results show that \ac{arden} with pruning achieves better performance for compensating frequency-dependent quadrature imbalance and power amplifier nonlinearity than other \ac{nn}-based and Volterra-based models, while requiring less or similar complexity.
\end{abstract}
\section{Introduction}
\Acf{rf} direct conversion transceivers suffer from multiple hardware impairments due to analog hardware imperfections~\cite{abidi1995direct} such as non-ideal \acfp{dac}, nonlinear active \acfp{lpf}, imperfect \acfp{lo}, and nonlinear \acfp{pa}. These impairments induce various signal distortions which degrade the quality of the transmitted signal, leading to reduced performance in terms of throughput~\cite{gustavsson2014impact}. These impairments can be mitigated separately by different algorithms, but separate optimization of each algorithm makse their combination not globally optimal. 

\ac{pa} nonlinearity is one of the major hardware impairments~\cite{cripps2006rf}. In the frequency domain, \ac{pa} nonlinearity materializes as in-band errors and out-of-band emissions due to intermodulation and harmonic products~\cite{pedro2003intermodulation}. \acp{pa} further exhibit memory effects during operation over large bandwidths~\cite{ku2003behavioral}, i.e., past input signals have nonlinear effects on the instantaneous output of the \ac{pa}. To linearize the \ac{pa}, it is customary to apply \ac{dpd}~\cite{kim2001digital}, which compensates for the signal distortion caused by the \ac{pa} nonlinearity, so that the cascade of the \ac{dpd} and the \ac{pa} is a linear system.
\Acf{iq} imbalance is another major impairment~\cite{cao2009q}, which commonly reflects as gain and phase mismatches, where the gain mismatch is introduced by the gain difference of \acp{dac} and \acp{lpf} between the \ac{I} and \ac{Q} branches, and the phase mismatch is caused by the \ac{lo} imperfection during up- and down-conversions. Similar to the \ac{pa}, the \ac{iq} imbalance introduces nonlinear distortions with memory effects due to the nonlinear  \acp{lpf} and \acp{dac}. 

Separate impairment mitigation of the \ac{pa} and \ac{iq} modulator has some shortcomings, as nonlinear mixing of the individual effects occurs. While some methods have been proposed to mitigate both impairments jointly, they suffer from either limited performance or high computational complexity~\cite{Lauri_PH_2010,wang2018augmented}. 

Several methods have been proposed to mitigate the \ac{iq} imbalance and \ac{pa} nonlinearity: Volterra series-based~\cite{cao2009q, ding2004MP, zhu2006dynamic, GMP_2006, Complexity_Ali_2010, Lauri_PH_2010}, \ac{nn}-based~\cite{liu2004dynamic, isaksson2005radial, rawat2009adaptive, mkadem2011physically,Mee_2012,wang2018augmented,jaraut2018composite,wu2020residual}, and others~\cite{anttila2008frequency,ding2008compensation}. The works~\cite{anttila2008frequency, ding2008compensation, cao2009q} only focus on \ac{iq} imbalance, while~\cite{raz1998baseband, ding2004MP, zhu2006dynamic, GMP_2006} propose simplified versions of Volterra series~\cite{eun1997new} focusing only on the \ac{pa} nonlinearity. Their performance is limited when both impairments occur~\cite{Lauri_PH_2010}.  Joint impairment mitigation of both the \ac{iq} modulator and \ac{pa} is investigated in~\cite{Lauri_PH_2010}, which extends the \ac{ph} method~\cite{ding2004MP} by the \ac{fir} \ac{iq} imbalance model so that the extended \ac{ph} allows to jointly mitigate both \ac{iq} modulator and \ac{pa} impairments. Its performance, however, is limited for highly nonlinear \acp{pa} and \ac{iq} modulators due to the simplification of the Volterra series and the linearity of \ac{fir} filters. All above mentioned Volterra-based models can improve performance by increasing the nonlinear order and memory length, but at the expense of an exponentially increasing complexity, which limits their utilization in practice~\cite{Complexity_Ali_2010}.

As an alternative to Volterra-based methods, \acp{nn} for \ac{iq}-\ac{pa} impairments mitigation are studied in~\cite{liu2004dynamic, isaksson2005radial, rawat2009adaptive, mkadem2011physically,Mee_2012,wang2018augmented,jaraut2018composite,wu2020residual}. Among them, the \ac{mlp} is mostly chosen due to easy deployment and training. Based on the \ac{mlp}, the \ac{rvtdnn} was proposed by Liu \textit{et al.}~\cite{liu2004dynamic} for \ac{pa} behavioral modeling. It allows to learn nonlinearities with memory effects by feeding real-valued I and Q components of the original complex-valued signal with time-delays. Various variants of the \ac{rvtdnn} have been later proposed~\cite{rawat2009adaptive,mkadem2011physically, Mee_2012, wang2018augmented, jaraut2018composite,wu2020residual}. The works~\cite{rawat2009adaptive,mkadem2011physically,wu2020residual} only focus on the \ac{pa} nonlinearity, while~\cite{Mee_2012,wang2018augmented} and~\cite{jaraut2018composite} consider both frequency-flat \ac{iq} imbalance and \ac{pa} nonlinearity in \ac{siso} and \ac{mimo} transmitters, respectively. Specifically, our recent work~\cite{wu2020residual} combines residual learning with \ac{rvtdnn}, which is demonstrated to improve performance for \ac{pa} nonlinearity mitigation as well as reduce complexity compared with other \ac{rvtdnn} variants. A similar performance improvement is also shown in~\cite{bajajsingle} for compensating nonlinearities of a fiber-optic link using residual learning \acp{nn}. None of these \ac{nn}-based models consider the mitigation of nonlinear frequency-dependent \ac{iq} imbalance, which is considerable in practice~\cite{cao2009q}. More importantly, the high-complexity problem of \ac{nn}-based models is not tackled excepts in our previous work~\cite{wu2020residual}, which limits their usages in practice. 

In this paper, we investigate the performance and complexity of impairment mitigation models for the direct conversion transceiver with multiple hardware impairments. Particularly, we consider the joint mitigation of nonlinear frequency-dependent \ac{iq} imbalance and \ac{pa} nonlinearity. Our contributions are summarized as follows:
\begin{itemize}
    \item We propose an attention residual learning \ac{nn} based on the \ac{rvtdnn}~\cite{liu2004dynamic}, referred to as \acf{arden}, to compensate for signal distortions caused by multiple hardware impairments, including \ac{pa} nonlinearity and nonlinear frequency-dependent \ac{iq} imbalance. Experimental results show that \ac{arden} yields better performance compared to state-of-the-art methods, while simultaneously exhibiting less complexity.
    \item We interpret the presence of an attention mechanism when learning the behavior of \ac{iq}-\ac{pa} system. We show that neurons in the first hidden layer of \ac{arden} fed by shorter lag input signals contribute more to the output with larger weight magnitudes, so these neurons deserve more attention.
    \item We propose and analyze a \ac{nn} connection pruning algorithm to reduce complexity. Unimportant neural connections, i.e., those with weights with small magnitude, are gradually removed during the pruning process. Results show that pruning allows \ac{arden} to achieve better mitigation performance with less complexity.
    \item We evaluate the mitigation performance of different methods for a large complexity range. Experimental results illustrate that \ac{arden} with proper pruning factor performs the best over all complexity levels.
\end{itemize}
This paper extends~\cite{wu2020residual} by generalizing to a multiple hardware impairments system including the \ac{pa} and \ac{iq} modulator. The weighted shortcut connections, attention mechanism, and pruning algorithm are novel.


\section{System Model}\label{section:sys_model} 
\begin{figure*}[t]
    \centering
    \begin{tikzpicture}[font=\footnotesize, >=stealth,nd/.style={draw,fill=blue!10,circle,inner sep=0pt,minimum size=5pt}, blk/.style={draw,fill=blue!10,minimum height=0.8cm,text width=1.5cm, text centered},triangle/.style = {fill=blue!20, regular polygon, regular polygon sides=3 },x=1.cm,y=1cm]
\tikzset{amplifier/.pic={
\draw [fill=blue!10](0,2.5)--(5,0)--(0,-2.5)--cycle node at (1.5,0) {\begin{tabular}{c} PA \\ $f_{\text{PA}}$ \end{tabular}};}}
\tikzset{dac_diag/.pic={
\draw [fill=blue!10](0,1.5)--(2.8,1.5)--(5,0)--(2.8,-1.5)--(0,-1.5)--cycle node at (1.8,0) {\begin{tabular}{c} DAC \end{tabular}};}}
\path (0,0)coordinate(in) node(dpd)[blk,right=0.8cm of in,fill=red!10]{DPD\\$f_{\text{DPD}}$} coordinate[right=0.8 of dpd](iq_sep){}
	coordinate[above=1.2 of iq_sep](iq_sep_I){}
	coordinate[below=1.2 of iq_sep](iq_sep_Q){}
	node[blk, right=0.5 of iq_sep_I,minimum width=.5cm, minimum height=0.8cm](real){Re$\{\star\}$}
	node[blk, right=0.5 of iq_sep_Q,minimum width=.5cm, minimum height=0.8cm](imag){Im$\{\star\}$}
	coordinate[right=1.1 of real](dac_in_I)
	(dac_in_I)pic[scale=.25,outer sep =0pt] {dac_diag}
	coordinate[right=1.25 of dac_in_I](dac_out_I){}
	
	coordinate[right=1.1 of imag](dac_in_Q)
	(dac_in_Q)pic[scale=.25,outer sep =0pt] {dac_diag}
	coordinate[right=1.25 of dac_in_Q](dac_out_Q){}

	node[blk,right=0.5 of dac_out_I, minimum width=.5cm, minimum height=0.8cm](lpf_in_I){LPF}
	node[blk,right=0.5 of dac_out_Q, minimum width=.5cm, minimum height=0.8cm](lpf_in_Q){LPF}
	node[nd,right=1.9 of lpf_in_I,minimum size=.5cm](mixer_I){$\times$}
	node[nd,right=1.9 of lpf_in_Q,minimum size=.5cm](mixer_Q){$\times$}
	node[nd,minimum size=.7cm](lo) at ($(mixer_I)!0.5!(mixer_Q)$){LO}
	coordinate[right=1.1 of mixer_I](pa_in_I){}
	coordinate[right=1.1 of mixer_Q](pa_in_Q){}
	node[nd](pa_in) at ($(pa_in_I)!0.5!(pa_in_Q)$){$+$}

	coordinate[right=1 of pa_in](pa)
	(pa)pic[scale=0.25,outer sep =0pt] {amplifier}
	coordinate[right=1.25 of pa](pa_out){}
	coordinate[right=.8 of pa_out](out){}
	node[blk,fill opacity=0, dashed, line width=1, minimum width=3.85cm,minimum height=1.1cm](h_i)at ($(dac_out_I)!0.37!(lpf_in_I)$){} 
	node[blk,fill opacity=0, dashed, line width=1, minimum width=3.85cm,minimum height=1.1cm](h_q)at ($(dac_out_Q)!0.37!(lpf_in_Q)$){}
	node[above=0 of h_i](){$f_{\text{I}}$}
	node[below=0 of h_q](){$f_{\text{Q}}$}
	;

\draw[->] (in)--node[above]{$u(n)$}(dpd);
\draw[-] (dpd)--node[above]{$x(n)$}(iq_sep);
\draw[-] (iq_sep)--node[above]{}(iq_sep_I);
\draw[-] (iq_sep)--node[above]{}(iq_sep_Q);
\draw[->] (iq_sep_I)--node[above]{}(real);
\draw[->] (iq_sep_Q)--node[above]{}(imag);

\draw[->] (real)--node[above]{$x_{\text{I}}(n)$}(dac_in_I);
\draw[-] (dac_out_I)--node[above]{}(lpf_in_I);
\draw[->] (imag)--node[below]{$x_{\text{Q}}(n)$}(dac_in_Q);
\draw[->] (dac_out_I)--node[below]{}(lpf_in_I);
\draw[->] (dac_out_Q)--node[below]{}(lpf_in_Q);
\draw[->] (lpf_in_I)--node[above]{$s_{\text{I}}(n)$}(mixer_I);
\draw[->] (lpf_in_Q)--node[below]{$s_{\text{Q}}(n)$}(mixer_Q);

\draw[->] (lo)--node[right]{$0^{\circ}$}(mixer_I);
\draw[->] (lo)--node[right]{$\frac{\pi}{2}+\phi$}(mixer_Q);
\draw[-] (mixer_I)--node[above]{$z_{\text{I}}(n)$}(pa_in_I);
\draw[-] (mixer_Q)--node[below]{$z_{\text{Q}}(n)$}(pa_in_Q);
\draw[->] (pa_in_I)--node[right]{}(pa_in);
\draw[->] (pa_in_Q)--node[right]{}(pa_in);
\draw[->] (pa_in)--node[above]{$z(n)$}(pa);
\draw[->] (pa_out)--node[above]{$y(n)$}(out);

\end{tikzpicture}
    \caption{Block diagram of the \ac{dpd}-\ac{iq}-\ac{pa} system. The \ac{dpd} block compensates for signal distortions caused by multiple hardware impairments in the direct conversion transmitter including non-ideal DACs, nonlinear LPFs, imperfect LO, and nonlinear PA.}
    \label{fig:IQI_PA_block_diagram}
\end{figure*}
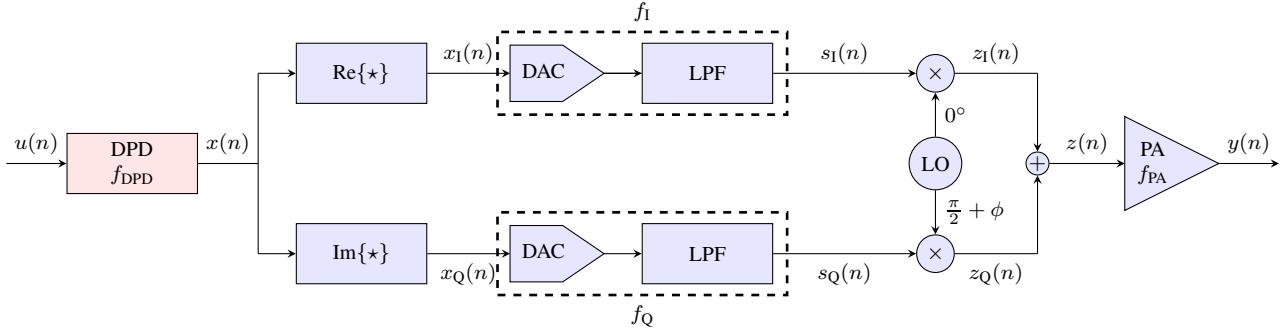
The block diagram of a direct conversion transmitter is shown in Fig.\,\ref{fig:IQI_PA_block_diagram}. The hardware impairments of the \acp{dac}, \acp{lpf}, \ac{lo}, and the \ac{pa} introduce \ac{iq} imbalance and \ac{pa} nonlinearity, which the \ac{dpd} placed before the hardware components tris to compensate. We now describe \ac{iq} imbalance, \ac{pa} nonlinearity, and \ac{dpd} in detail.
\subsection{\ac{iq} Imbalance}
As shown in Fig.\,\ref{fig:IQI_PA_block_diagram}, considering a discrete-time baseband signal $x(n)$ to be modulated by the \ac{iq} modulator, its real and imaginal parts, $x_{\text{I}}(n)$ and $x_{\text{Q}}(n)$ are sent to the I and Q branches of the modulator, respectively. We consider both wideband and frequency-dependent \ac{iq} imbalances. The wideband \ac{iq} imbalance is due to memoryless nonlinearities of non-ideal \acp{dac} caused by quantization noise and clipping, while the frequency-dependent \ac{iq} imbalance is due to nonlinearities with memory effects of imperfect and non-equal \acp{lpf}. The combination of \ac{dac} and \ac{lpf} is represented by the nonlinear function $f_{\text{I}}:\mathbb{R}^{L_1+1}\rightarrow \mathbb{R}$ and $f_{\text{Q}}:\mathbb{R}^{L_1+1}\rightarrow \mathbb{R}$ for the \ac{I} and \ac{Q} branches, respectively, where $L_1$ is the memory length. Denote the output of the \ac{dac}-\ac{lpf} for the I and Q branches as $s_{\text{I}}(n)$ and $s_{\text{Q}}(n)$, respectively. Their input-output relations can be expressed as
\begin{align}
s_{\text{I}}(n) &= f_{\text{I}}(x_{\text{I}}(n),\ldots,x_{\text{\text{I}}}(n-L_1)) = f_{\text{I}}(\boldsymbol{x}_{\text{I}}^{L_1}), \\
s_{\text{Q}}(n) &= f_{\text{Q}}(x_{\text{Q}}(n),\ldots,x_{\text{\text{Q}}}(n-L_1)) = f_{\text{Q}}(\boldsymbol{x}_{\text{Q}}^{L_1}),
\end{align}
where $\boldsymbol{x}_{\text{I}}^{L_1}=[x_{\text{I}}(n),\ldots,x_{\text{I}}(n-L_1)]^{\mathsf{T}}$, and  $\boldsymbol{x}_{\text{Q}}^{L_1}=[x_{\text{Q}}(n),\ldots,x_{\text{Q}}(n-L_1)]^{\mathsf{T}}$.

The \ac{dac}-\ac{lpf} outputs are up-converted by mixers, where a phase imbalance $\phi$ is introduced, caused by \ac{lo} imperfection. The output of the \ac{iq} modulator is
\begin{equation}
    z(n) = z_{\text{I}}(n) + \jmath z_{\text{Q}}(n),
    \label{eq:iqi_z_1}
\end{equation}
where $z_{\text{I}}(n)=s_{\text{I}}(n)-\sin (\phi) s_{\text{Q}}(n)$ and $z_{\text{Q}}(n)=\cos (\phi) s_{\text{Q}}(n)$. Equation~\eqref{eq:iqi_z_1} can be rewritten as
\begin{equation}
\begin{aligned}
    z(n) &= s_{\text{I}}(n)-\sin (\phi) s_{\text{Q}}(n) + \jmath \cos (\phi) s_{\text{Q}}(n)\\
    &= s_{\text{I}}(n) +\jmath e^{\jmath\phi}s_{\text{Q}}(n) \\
    &= f_{\text{I}}(\boldsymbol{x}_{\text{I}}^{L_1}) + \jmath e^{\jmath\phi} f_{\text{Q}}(\boldsymbol{x}_{\text{Q}}^{L_1}),\label{eq:z_time_domain_3}
\end{aligned}
\end{equation}
Due to the difference between \acp{dac} and \acp{lpf} of the I and Q branches, $f_{\text{I}}$ and $f_{\text{Q}}$ present different nonlinearities and memory effects, which leads to \ac{iq} imbalances with both frequency-independent and frequency-dependent components. For ease of notation~\eqref{eq:z_time_domain_3}, we use a single function $f_{\text{IQ}}:\mathbb{C}^{L_1+1}\rightarrow \mathbb{C}$ with memory length $L_1$ to represent the \ac{iq} modulator system, so \eqref{eq:z_time_domain_3} can be rewritten as
\begin{equation}
    z(n) = f_{\text{IQ}}(x(n),\ldots,x(n-L_1)) = f_{\text{IQ}}(\boldsymbol{x}^{L_1}),
    \label{eq:z_func}
\end{equation}
where $\boldsymbol{x}^{L_1}=[x(n),\ldots,x(n-L_1)]^{\mathsf{T}}$. Note that for an ideal \ac{iq} modulator $\phi=0$, $L_1=0$, and $z(n)=x(n)$.

\subsection{\ac{pa} Nonlinearity}
The modulated signal $z(n)$ is amplified by the \ac{pa}, which behaves as a nonlinear system with memory effects, i.e., the \ac{pa} output at any time instant depends on the current instantaneous input and previous inputs. Memory effects are mainly due to the frequency-dependent behavior of the \ac{pa} and thus more considerable for wideband signals. We define the PA as a function $f_{\text{PA}}:\mathbb{C}^{L_2+1}\rightarrow \mathbb{C}$ with input $z(n)$ and output $y(n)$, and memory length $L_2$,
\begin{equation}
    y(n) = f_{\text{PA}}(z(n),\ldots,z(n-L_2)) = f_{\text{PA}}(\boldsymbol{z}^{L_2}),
    \label{eq:pa_in_out_relation}
\end{equation}
where $\boldsymbol{z}^{L_2}=[z(n),\ldots,z(n-L_2)]^{\mathsf{T}}$. For an ideal \ac{pa}, $L_2=0$ and $y(n)=G z(n)$, $G$ being the \ac{pa} gain.
\subsection{Digital Predistortion}
The \ac{dpd} is represented by the function $f_{\text{DPD}}:\mathbb{C}^{L_3+1}\rightarrow \mathbb{C}$ with memory length $L_3$ and input signal $u(n)$,
\begin{equation}
    x(n) = f_{\text{DPD}}(u(n),\ldots,u(n-L_3)) = f_{\text{DPD}}(\boldsymbol{u}^{L_3}),
    \label{eq:dpd_in_out_relation}
\end{equation}
where $\boldsymbol{u}^{L_3}=[u(n),\ldots,u(n-L_3)]^{\mathsf{T}}$. 

Substituting~\eqref{eq:z_func} into~\eqref{eq:pa_in_out_relation}, we can rewrite $y(n)$ as
\begin{equation}
\begin{aligned}
    y(n)=& f_{\text{PA}}(f_{\text{IQ}}(\boldsymbol{x}^{L_1}),\ldots,f_{\text{IQ}}(\boldsymbol{x}_{n-L_2}^{L_1}))
    \\
    =& f_{\text{IQ-PA}}(x(n),\ldots, x(n-L_1-L_2)) \\
    =& f_{\text{IQ-PA}}(\boldsymbol{x}^{L_1+L_2}) ,
    \label{eq:cascade_iq_pa_3}
    \end{aligned}
    \end{equation}
where the function $f_{\text{IQ-PA}}: \mathbb{C}^{L_1+L_2+1}\rightarrow \mathbb{C}$ represents the \ac{iq}-\ac{pa} system. The system resulting from the cascade of the \ac{iq} modulator and the \ac{pa} has memory length $(L_1+L_2)$.  

The input-output relation of the whole system is obtained by substituting~\eqref{eq:dpd_in_out_relation} into~\eqref{eq:cascade_iq_pa_3} as
\begin{equation}
\begin{aligned}
y(n)= &f_{\text{IQ-PA}}(f_{\text{DPD}}(\boldsymbol{u}_{n}^{L_3}),\ldots,f_{\text{DPD}}(\boldsymbol{u}_{n-L_1-L_2}^{L_3}))
 \\
= & f_{\text{DPD-IQ-PA}}(\boldsymbol{u}^{L_1+L_2+L_3}),
\label{eq:cascade_dpd_iq_pa_1}
\end{aligned}
\end{equation}
where the function $f_{\text{DPD-IQ-PA}} \in \mathbb{C}^{L_1+L_2+L_3}\rightarrow \mathbb{C}$ denotes the \ac{dpd}-\ac{iq}-\ac{pa} system with memory length $(L_1+L_2+L_3)$.

Ideally, the \ac{dpd} would make the cascade \ac{dpd}-\ac{iq}-\ac{pa} linear, in which case \eqref{eq:cascade_dpd_iq_pa_1} would reduce to the linear function $y(n)=Gu(n)$. Unfortunately, this is infeasible in practice due to the presence of hardware impairments such as \ac{pa} clipping and thermal noise, which can not be compensated for. \ac{dpd} methods aim, therefore, to make the \ac{dpd}-\ac{iq}-\ac{pa} system as linear as possible by minimizing the \ac{mse} between the \ac{pa} output $y(n)$ and \ac{dpd} input $u(n)$,
\begin{equation}
    \hat{f}_{\text{DPD}} = \text{arg }\underset{f_{\text{DPD}}}{\text{min }}\mathbb{E}[|f_{\text{DPD-IQ-PA}}(\boldsymbol{u}^{L_1+L_2+L_3})-u(n)|^2],
    \label{eq:arg_dpd_direct}
\end{equation}
where $\mathbb{E}[\cdot]$ denotes expectation.

\section{Preliminaries}
\subsection{\ac{dpd}-parameter Identification by ILA}\label{section:ILA}
In practice, estimating the parameters of the \ac{dpd} function $f_{\text{DPD}}$ through \eqref{eq:arg_dpd_direct} is troublesome as the \ac{iq}-\ac{pa} system is generally a combination of black boxes, i.e., unknown $f_{\text{IQ-PA}}$. The \ac{dla}~\cite{dla_Zhou_2006} solves this problem by approximating $f_{\text{IQ-PA}}$ as a differential model, which allows to iteratively identify \ac{dpd} parameters through a gradient-based method. However, the accuracy of the identified \ac{dpd} is seriously affected by the accuracy of the approximated $f_{\text{IQ-PA}}$, and the identification process of \ac{dla} is highly complex due to numerous updating iterations.

Instead, the \ac{ila}~\cite{eun1997new} indirectly estimates \ac{dpd} parameters by learning the inverse behavior of the \ac{iq}-\ac{pa} system, i.e., $f_{\text{IQ-PA}}^{-1}$, referred to as the \textit{post-distorter}, which is then used as the \textit{pre-distorter} for \ac{dpd}~\cite{schetzen1976theory}. Thus, the \ac{dpd} estimator \eqref{eq:arg_dpd_direct} using \ac{ila} is changed to
\begin{equation}
    \hat{f}_{\text{DPD}} = \text{arg }\underset{f_{\text{IQ-PA}}^{-1}}{\text{min }}\mathbb{E}[|f_{\text{IQ-PA}}^{-1}(\boldsymbol{y}^{L_1+L_2}) - x(n)|^2],
    \label{eq:arg_dpd_ila}
\end{equation}
where $\boldsymbol{y}^{L_1+L_2}=[y(n),...,y(n-L_1-L_2)]$. \ac{ila} is the most used identiﬁcation method due to simple implementation and excellent performance~\cite{Jessica_ILA_Var}. Therefore, we consider ILA as the identiﬁcation method for DPD in this paper.
\subsection{Attention Residual Learning}\label{section:attention_mecha}
The attention mechanism has been widely used in many areas such as machine translation~\cite{luong2015effective}, and image classification~\cite{wang2017residual}. Based on the application or prior knowledge of the learning object, important features of a learning objective are highlighted by artificial attentions,  such as the shape of an image or a specific word in a sentence, and those attentions help in the learning process.

Consider an unknown system $f$ with input $x$ and output $y$, i.e., $y=f(x)$. With some prior knowledge of this system, a prior estimation $f_{\text{prior}}$ of $f$ can be made, which can be helpful to further find a more accurate estimation of $f$. Thus, we say that $f_{\text{prior}}$ \textit{deserves} more attention when learning $f$. This attention mechanism can be implemented by extracting $f_{\text{prior}}$ from $f$ as
\begin{equation}
    y = \underbrace{f_{\text{prior}}(x)}_{=f_{\text{atten}}(x)} + \underbrace{f(x) -f_{\text{prior}}(x)}_{=f_{\text{resid}}(x)},
    \label{eq:sys_atten_residual}
\end{equation}
where we refer to the extracted component $f_{\text{prior}}(x)$ as the \textit{attentive function}, denoted by $f_{\text{atten}}$, and the residual component $f(x) - f_{\text{prior}}(x)$ as the \textit{residual function}, denoted by $f_{\text{resid}}(x)$. Thus, the prior function $f_{\text{prior}}(x)$ is now considered explicitly, i.e., being attentive, during the learning of $f(x)$, which helps in the learning process. In some scenarios such as image recognition, no prior knowledge of $f$ is given, so $f_{\text{prior}}(x)$ is set to $x$, which makes \eqref{eq:sys_atten_residual} reduce to residual learning~\cite{ResNet_Kaiming}. They authors in~\cite{ResNet_Kaiming} have shown that learning a residual function is more effective than learning its corresponding original function. 

\section{Attention Residual Learning \\ Neural Network} \label{section:arden}
In this section, we introduce the proposed \ac{arden} to mitigate impairments of the \ac{iq}-\ac{pa} system and \ac{nn} pruning to reduce complexity.

\subsection{Attention Residual Learning for \ac{iq}-\ac{pa} System}\label{section:arl_iq_pa}
Since the inverse behavior of the \ac{iq}-\ac{pa} system contains the same type of hardware impairments as its forward behavior, learning the forward or backward behaviors reduces to changing the input-output and vice versa. Thus, for ease of understanding, we apply the attention mechanism described in Section~\ref{section:attention_mecha} to analyze the forward behavior of the \ac{iq}-\ac{pa} system.

For the \ac{pa}-\ac{iq} system, the  system function $f$ in \eqref{eq:sys_atten_residual} corresponds to~\eqref{eq:cascade_iq_pa_3}, and the prior function $f_{\text{prior}}$ can be obtained by ignoring nonlinearities and memory effects, in which case, $f_{\text{IQ-PA}}$ reduces to a linear function,
\begin{equation}
    y(n) = W x(n)
    \label{eq:iq-pa-linear}
\end{equation}
where $W$ is a mix of the linear narrowband \ac{iq} imbalance and \ac{pa} gain $G$. Following~\eqref{eq:sys_atten_residual}, we can extract the linear component \eqref{eq:iq-pa-linear} from the original function \eqref{eq:cascade_iq_pa_3} as 
\begin{equation}
    y(n) = \underbrace{W x(n)}_{f_{\text{atten}}^{\text{IQ-PA}}(x(n))} + \underbrace{f_{\text{IQ-PA}}(\boldsymbol{x}^{L_1+L_2}) - W x(n)}_{=f_{\text{res}}^{\text{IQ-PA}}(\boldsymbol{x}^{L_1+L_2})},
    \label{eq:iq-pa-residual}
\end{equation}
where $f_{\text{atten}}^{\text{IQ-PA}}$ and $f_{\text{resid}}^{\text{IQ-PA}}$ denote the attentive and residual functions of the \ac{iq}-\ac{pa} system, respectively. Thus, learning the original unknown \ac{iq}-\ac{pa} behavior $f_{\text{IQ-PA}}$ reduces to learning the residual nonlinear behavior, as the linear behavior is always activated as attentions. Here, we refer to $W$ as the \textit{attention weight} as it decides how much attention we pay to the linear input-output relation.

We remark that our attention residual learning is different from the residual learning in~\cite{ResNet_Kaiming}, as we only extract a specific part of the input, i.e., the current input signal $x(n)$, instead of the whole input sequence. Also, here the residual function $f_{\text{res}}^{\text{IQ-PA}}$ has a practical meaning that represents the nonlinear input-output relation in the \ac{iq}-\ac{pa} system, whereas the residual function in~\cite{ResNet_Kaiming} does not. Furthermore, while our method is based on the attention mechanism, it also differs from the attention models in~\cite{luong2015effective,wang2017residual}. Particularly, the focus of our attention is always fixed on the current input signal, i.e., the attention weights for other input signals are always zero. The selection of attention weights is based on the prior knowledge of the \ac{iq}-\ac{pa} system, as there is a strong linear relation between the input and output signals.

\subsection{\ac{arden} Architecture}\label{section:archit_arden}
\begin{figure}[t]
    \centering
    \begin{tikzpicture}
[font=\footnotesize, draw=black!50, cnode/.style={draw=black!50,fill=black!25,minimum width=4.5mm,circle}, x=0.55cm,y=0.55cm
]

    \tikzstyle{annot} = [text centered]
    \node at (0,-1.7) {$\vdots$};
    \node at (3,-2) {$\vdots$};
    \node at (5,-2) {$\vdots$};
    \node at (4,0) {$\dots$};
    \node at (4,-1) {$\dots$};
	\node at (4,-2.2) {$\dots$};
	\node at (4,-4) {$\dots$};
    \node[cnode=blue,fill=red!30,pin={[pin edge={<-, line width=0.6}]left:$s^{\text{I}}_{\text{in}}(n)$}] (s11) at (0,2) {};
    \node[cnode=blue,,fill=red!30,pin={[pin edge={<-, line width=0.6}]left:$s^{\text{Q}}_{\text{in}}(n)$}] (s12) at (0,1) {};
    \node[cnode=gray,fill=green!30,pin={[pin edge={<-, line width=0.6}]left:$s^{\text{I}}_{\text{in}}(n-1)$}] (s13) at (0,0) {};
    \node[cnode=gray,fill=green!30,pin={[pin edge={<-, line width=0.6}]left:$s^{\text{Q}}_{\text{in}}(n-1)$}] (s14) at (0,-1) {};
    \node[cnode=gray,fill=green!30,pin={[pin edge={<-, line width=0.6}]left:$s^{\text{Q}}_{\text{in}}(n-M)$}] (s15) at (0,-3) {};
	\node[cnode=gray,fill=green!30,pin={[pin edge={<-, line width=0.6}]left:$s^{\text{Q}}_{\text{in}}(n-M)$}] (s16) at (0,-4) {};	
    \node[cnode=gray,fill=blue!30,label=180:] (s21) at (3,0) {};
    \node[cnode=gray,fill=blue!30,label=180:] (s22) at (3,-1) {};
    \node[cnode=gray,fill=blue!30,label=180:] (s23) at (3,-4) {};
    \draw[solid] (s11) -- (s21);
    \draw[dotted] (s11) -- (s22);
    \draw[solid] (s11) -- (s23);
    \draw[dotted] (s12) -- (s21);
    \draw[solid] (s12) -- (s22);
    \draw[dotted] (s12) -- (s23);
    \draw[solid] (s13) -- (s21);
    \draw[dotted] (s13) -- (s22);
    \draw[solid] (s13) -- (s23);
    \draw[dotted] (s14) -- (s21);
    \draw[solid] (s14) -- (s22);
    \draw[dotted] (s14) -- (s23);
    \draw[dotted] (s15) -- (s21);
    \draw[solid] (s15) -- (s22);
    \draw[dotted] (s15) -- (s23);
    \draw[dotted] (s16) -- (s21);
    \draw[solid] (s16) -- (s22);
    \draw[dotted] (s16) -- (s23);
    
    \node[cnode=gray,fill=blue!30,label=180:] (s31) at (5,0) {};
    \node[cnode=gray,fill=blue!30,label=180:] (s32) at (5,-1) {};
	\node[cnode=gray,fill=blue!30,label=180:] (s33) at (5,-4) {};
	\node[cnode=blue,fill=red!30,pin={[pin edge={->, line width=0.6}]right:$\hat{s}^{\text{I}}_{\text{out}}(n)$}] (s41) at (8,2) {};
    \node[cnode=blue,fill=red!30,pin={[pin edge={->, line width=0.6}]right:$\hat{s}^{\text{Q}}_{\text{out}}(n)$}] (s42) at (8,1) {};
    \draw[solid] (s31) -- (s41);
    \draw[dotted] (s31) -- (s42);
    \draw[dotted] (s32) -- (s41);
    \draw[solid] (s32) -- (s42);
    \draw[solid] (s33) -- (s41);
    \draw[dotted] (s33) -- (s42);    
    
    \foreach \x in {1,...,2}
    {   \foreach \y in {1,...,2}
        {   \draw[color=red!50,, line width=1] (s1\x) -- (s4\y);
        }
    }	
   	\node[annot, below of=s16](l1) {\begin{tabular}{c} Input layer\\ $\in\mathbb{R}^{2(M+1)}$ \end{tabular}};
   	\node[annot](l2) at (4,-5.82) {\begin{tabular}{c} Hidden layers\\ $\in\mathbb{R}^{D_k}$ \end{tabular}};
   	\node[annot](l3) at (8,-5.82) {\begin{tabular}{c} Output layer \\ $\in\mathbb{R}^{2}$ \end{tabular}};
   	\node[annot, red!70](l3) at (3.7,2.6) {\begin{tabular}{c} Attention \\ Shortcuts \end{tabular}};
    \end{tikzpicture}
    \caption{Architecture of the proposed \ac{arden} with arbitrary connections being pruned. Dotted and solid lines between neurons represents pruned and remained connections. Fed by the real-valued I and Q components of the current and historical time instant signals, \ac{arden} returns estimations of the real-valued I and Q components of the current time instant output signal. When using ILA to estimate DPD parameters, $s_{\text{in}}(n)=y(n)$ and $s_{\text{out}}(n)=x(n)$.}
    \label{fig:archit_arden}
\end{figure}
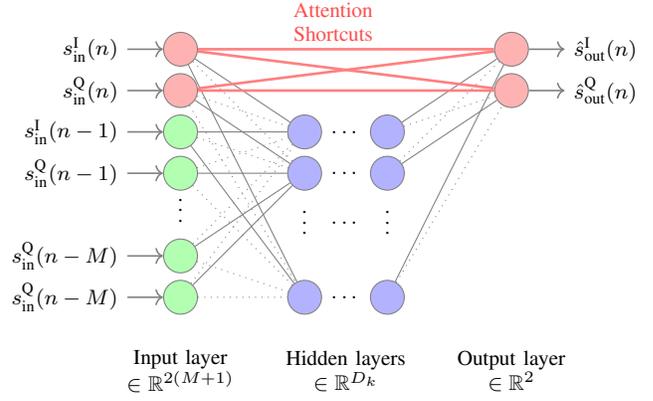
Based on the \ac{mlp}, we propose a novel \ac{nn} by considering the proposed residual learning attention method~\eqref{eq:iq-pa-residual} for the \ac{iq}-\ac{pa} system, referred to as \ac{arden}, and Fig.\,\ref{fig:archit_arden} shows the block diagram of \ac{arden} with arbitrary connections pruned by the NN pruning algorithm (dotted lines), which is described in Section~\ref{section:prun_alg}. \ac{arden} consists of $K$ fully connected layers with $(K-2)$ hidden layers. The number of neurons in layer $k$ is denoted by $D_k$. The input vector of layer $k$ is denoted by $\boldsymbol{s}_k \in \mathbb{R}^{D_{k-1}}$ for $k>1$. The input and output vectors of the input and output layers are $\boldsymbol{s}_1$ and $\boldsymbol{s}_{K+1}$.~\footnote{During the DPD parameter estimation using ILA, $\boldsymbol{s}_1$ and $\boldsymbol{s}_{K+1}$ are formed by $y(n)$ and $x(n)$, respectively, and vice versa when deploy ARDEN as DPD.} Define a complex-valued signal with sample $s_{\text{in}}(n) = s_{\text{in}}^{\text{I}}(n) + \jmath s_{\text{in}}^{\text{Q}}(n)$ at time instant $n$ as the input of \ac{arden}. The real-valued input vector $\boldsymbol{s}_1$ is formed by concatenating the current and previous time instants of the input signal,
\begin{equation}
\begin{aligned}
\boldsymbol{s}_1 = [&s_{\text{in}}^{\text{I}}(n), s_{\text{in}}^{\text{Q}}(n),\ldots,s_{\text{in}}^{\text{I}}(n-M), s_{\text{in}}^{\text{Q}}(n-M)]^{\mathsf{T}},
    \label{eq:r2tdnn_s_1}
\end{aligned}
\end{equation}
where $M$ denotes the number of time delays for the input signal. The time-delayed inputs allow \ac{arden} to capture the memory effects of the transmitter, and separating the real-valued signals allows the use of a simple real-valued training algorithm. In total, the number of neurons for the input layer is $D_1 = (2M + 2)$.

We denote the weight matrix that connects layer $k-1$ and $k$ by $\boldsymbol{W}_k \in \mathbb{R}^{D_k \times D_{k-1}}$, for $k>1$, the $j$th column weight vector of $\boldsymbol{W}_k$ by $[\boldsymbol{W}_k]_j \in \mathbb{R}^{D_k}$, and the corresponding bias vector by $\boldsymbol{b}_k \in \mathbb{R}^{D_k}$. For $k>1$, layers $k-1$ and $k$ are fully connected as
\begin{equation}
\boldsymbol{s}_{k+1} = \sigma (\boldsymbol{W}_{k} \boldsymbol{s}_{k} +\boldsymbol{b}_{k}),
    \label{eq:r2tdnn_fully_connect}
\end{equation}
where $\sigma$ denotes the element-wise activation function. To output a full range of values, the output layer is a linear layer, i.e., function $\sigma$ for the output layer is an identity mapping function, with number of neurons $D_K=2$ corresponding to the I and Q output signals. 

The attentive function in~\eqref{eq:iq-pa-residual} is implemented in \ac{arden} by weighted \textit{shortcut connections} between the instantaneous input and output signals as 
\begin{equation}
    f_{\text{atten}}^{\text{IQ-PA}} = \boldsymbol{W}_a [s_{\text{in}}^{\text{I}}(n), s_{\text{in}}^{\text{Q}}(n)]^\mathsf{T}
    \label{eq:arden_f_atten},
\end{equation}
where the trainable weight matrix $\boldsymbol{W}_a \in \mathbb{R}^{2\times 2}$ corresponds to a complex-valued $W$ in~\eqref{eq:iq-pa-linear}. Assuming no prior knowledge of the linear \ac{iq} imbalance and \ac{pa} gain,  we initialize $\boldsymbol{W}_a$ as a $2\times 2$ identity matrix. The shortcut connections are shown by red lines in Fig.\,\ref{fig:archit_arden}. The residual function in~\eqref{eq:iq-pa-residual} is implemented by the hidden layers of \ac{arden}. Thus, the output of the output layer can be expressed as
\begin{equation}
    \hat{\boldsymbol{s}}_{\text{out}} \triangleq \boldsymbol{s}_{K+1} = \underbrace{\boldsymbol{W}_a [s_{\text{in}}^{\text{I}}(n), s_{\text{in}}^{\text{Q}}(n)]^\mathsf{T}}_{f_{\text{atten}}^{\text{IQ-PA}}} + \underbrace{\boldsymbol{W}_K \boldsymbol{s}_{K} + \boldsymbol{b}_K}_{f_{\text{res}}^{\text{IQ-PA}}},
    \label{eq:arden_S_out}
\end{equation}
where $\hat{\boldsymbol{s}}_{\text{out}}\in \mathbb{R}^{2}$ consists of the I and Q output signal estimations $\hat{s}_{\text{out}}^{\text{I}}(n)$ and $\hat{s}_{\text{out}}^{\text{Q}}(n)$ of the complex-valued output signal $s_{\text{out}}(n)$ at time instant $n$, respectively.

Denote all weight matrices and bias vectors as $\boldsymbol{W}=\{\boldsymbol{W}_1, \ldots, \boldsymbol{W}_K, \boldsymbol{W}_a\}$ and $\boldsymbol{b}=\{\boldsymbol{b}_1, \ldots, \boldsymbol{b}_K\}$.  $\boldsymbol{W}$ and $\boldsymbol{b}$ can be learned through gradient descent by minimizing the \ac{mse} between the estimation $\hat{\boldsymbol{s}}_{\text{out}}$ and observation $\boldsymbol{s}_{\text{out}}$,
\begin{equation}
    (\boldsymbol{W}^{\*},\boldsymbol{b}^{\*}) = \arg \underset{\boldsymbol{W},\boldsymbol{b}}{\min} \mathbb{E}[|\boldsymbol{s}_{\text{out}}-\hat{\boldsymbol{s}}_{\text{out}}|^2]. 
\end{equation}

\subsection{Neural Network Pruning}\label{section:prun_alg}
\begin{algorithm}[t]
\caption{: Magnitude-based pruning for layer $k$.}
 \hspace*{\algorithmicindent} \textbf{Input:} Total training step $N$. Pruning interval $\Delta N$.
 
\begin{algorithmic}[1]
\FOR{$n = 1 \rightarrow N$}
\IF{$n/\Delta N = $ integer}
    \STATE Calculate $\eta_n$ using \eqref{eq:prune_eta}
    \STATE Calculate $N_{\text{p}}$ using \eqref{eq:prune_Np}
    \STATE Zero $N_{\text{p}}$ weights of smaller magnitude in $\boldsymbol{W_k}$ 
    \STATE Zero the corresponding $N_{\text{p}}$ masks in $\boldsymbol{M_k}$
\ELSE
\STATE Update weights in $\boldsymbol{W}_k$ with non-zero masks via back-propagation
\ENDIF
\ENDFOR 
\STATE Remove $\boldsymbol{M}_k$
 \end{algorithmic}
  \label{alg:pruning}
 \end{algorithm}
\acp{nn} have been shown to achieve good performance in many tasks. However, the high computation complexity makes the deployment of \acp{nn} challenging in resource-constrained scenarios where the resource overhead for each chain, and thus for each \ac{dpd}, is limited. Hence, it is crucial to reduce the complexity of \acp{nn} for \ac{dpd}.

To reduce the complexity requirement, one popular technique that has been studied in recent years is \ac{nn} pruning~\cite{han2015deep, zhu2017prune}, which reduces the \ac{nn} size by removing unimportant neurons and/or connections. We apply the pruning method in~\cite{zhu2017prune} to reduce the complexity of \ac{arden}. Pruning works on each layer by adding a binary mask with the same size as the layer's weight matrix, in which a zero indicates that the weight is pruned. Let $\boldsymbol{M}_k \in \mathbb{R}^{D_k \times D_{k-1}}$ denote the binary mask matrix of layer $k$. The connection between layer $(k-1)$ and $k$ in~\eqref{eq:r2tdnn_fully_connect} with pruning can be rewritten as
\begin{equation}
\boldsymbol{s}_{k+1} = \sigma((\boldsymbol{M}_k\odot\boldsymbol{W}_{k}) \boldsymbol{s}_{k} +\boldsymbol{b}_{k}),
    \label{eq:prune_fully_connect}
\end{equation}
where $\odot$ denotes the Hadamard product operator. Note that shortcut connections in \ac{arden} are not pruned so as to keep the attention function $f_{\text{atten}}^{\text{IQ-PA}}$ always active.

Define the NN sparsity $\eta$ as the ratio of the number of zero weights to the total number of weights. Given a total number of training steps $N$, weights are pruned every $\Delta N$ steps, referred to as \textit{pruning process}. Denote $\eta$ at step $n$ as $\eta_n$, which is gradually increased to the desired sparsity $\eta_{\text{d}}$ by~\cite{zhu2017prune}
\begin{equation}
\eta_n = \eta_{\text{d}} -\eta_{\text{d}}\left(1-\frac{\lf n/\Delta N\rf}{N} \right)^3.
\label{eq:prune_eta}
\end{equation} 
The intuition behind \eqref{eq:prune_eta} is to prune rapidly at the beginning and gradually prune less weights when the sparsity grows high. After each pruning step, non-pruned weights are retrained for $N-1$ training steps, referred to as \textit{retraining process}, to alleviate the loss caused by pruning.

The pruning for layer $k$ of \ac{arden} is in Algorithm~\ref{alg:pruning}. During each pruning step, $\eta_n$ is calculated using \eqref{eq:prune_eta}. To meet $\eta_n$ for the layer with a total number of weights $N_{\text{w}}$, the number of weights $N_\text{p}$ needed to be pruned is calculated by
\begin{equation}
    N_\text{p} = N_{\text{w}} \times(1-\eta_n).
    \label{eq:prune_Np}
\end{equation}
Then, weights in $\boldsymbol{W}_k$ are sorted by magnitudes, and the $N_{\text{p}}$ weights of smaller magnitude are masked to zero by setting the corresponding values in $\boldsymbol{M}_k$ to zero. During each retraining step, weights in $\boldsymbol{W}_k$ with non-zero masks are updated through $N-1$ back-propagation steps. Once pruning is done, $\boldsymbol{M}_k$ is removed.

\subsection{Computational Complexity}\label{section:complexity_cal}
We focus on the \textit{running complexity}~\cite{Complexity_Ali_2010} of the DPD, which is defined as the number of calculations required for the inference of each output sample. Unlike the \textit{identification complexity} for estimating DPD parameters that is usually done off-line, the running complexity is a real-time cost, which heavily limits the system overhead. It can be quantified by the number of multiplications and additions operated, where each real-valued multiplication or addition accounts for one \ac{flop}~\cite[Table. I]{Complexity_Ali_2010}.

We measure the complexity of \ac{arden} in terms of the number of \acp{flop} as
\begin{equation}
    C_{\text{arden}} = 2 (1-\eta_{\text{d}}) \sum_{k=1}^{K-1}D_k D_{k+1} + 8,
    \label{eq:complexity_r2tdnn}
\end{equation}
Note that given a fixed size \ac{arden}, its complexity decreases linearly as $\eta_{\text{d}}$ increases.
where the factor $8$ corresponds to the number of \acp{flop} introduced by the shortcut connection. Note that the required number of \acp{flop} decreases nearly linearly with the desired network sparsity. The computational complexity of other \ac{rvtdnn}-based methods can be calculated in a similar way. 
\section{Experimental results}\label{section:results}

\subsection{Setup}
\subsubsection{Measurement Setup}
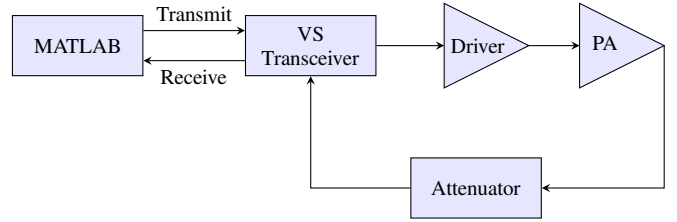
\begin{figure}[t]
    \centering
    \begin{tikzpicture}[font=\footnotesize, >=stealth,nd/.style={draw,fill=blue!10,circle,inner sep=0pt,minimum size=5pt}, blk/.style={draw,fill=blue!10,minimum height=0.8cm,text width=1.5cm, text centered}, x=0.9cm, y=0.9cm]
\tikzset{driver/.pic={
\draw [fill=blue!10](0,2.5)--(5,0)--(0,-2.5)--cycle node at (2,0) {Driver};}}
\tikzset{amplifier/.pic={
\draw [fill=blue!10](0,2.5)--(5,0)--(0,-2.5)--cycle node at (1.5,0) {\begin{tabular}{c} PA \end{tabular}};}}

\path (0,0)node(matlab)[blk]{MATLAB}  
	node(vst)[blk, right=1.5 of matlab]{VS\\Transceiver} 
	coordinate[below=0 of vst](vst_mid)
	coordinate[right=1 of vst](driv_in)
	(driv_in)pic[scale=0.25,outer sep =0pt]{driver}
	coordinate[right=1.25 of driv_in](driv_out)
	coordinate[right=2 of driv_in](pa_in)
	(pa_in)pic[scale=0.25,outer sep =0pt]{amplifier}
	coordinate[right=1.25 of pa_in](pa_out)
	node(att)[blk, below right=1.2 and 0.5 of vst]{Attenuator}

;

\draw[->] ([yshift=0.2 cm]matlab.east)--node[above]{Transmit}([yshift=0.2 cm]vst.west);
\draw[<-] ([yshift=-0.2 cm]matlab.east)--node[below]{Receive}([yshift=-0.2 cm]vst.west);

\draw[->] (vst)--(driv_in);
\draw[->] (driv_out)--(pa_in);
\draw[->] (pa_out)|-(att);
\draw[->] (att.west)-|(vst.south);

\end{tikzpicture}
    \caption{Block diagram of the RF WebLab. Digital signals are transmitted and received by the block of MATLAB.}
    \label{fig:weblab_block_diagram}
\end{figure}

The measurement setup is based on the RF WebLab~\cite{landin2015weblab}, which can be remotely accessed at \url{www.dpdcompetition.com}. Its block diagram is shown in Fig.~\ref{fig:weblab_block_diagram}. The block MATLAB includes all digital signal processing steps such as the \ac{dpd} identification, \ac{dpd} deployment, and artificial \ac{iq} imbalance generation. In the transmission stage, digital signals generated by MATLAB are converted into analog signals by a \ac{vst} PXIe-5646R VST, and then transmitted to the Gallium Nitride PA DUT (Cree CGH4006-TB) with a $40$ dB linear driver. In the receiving stage, through a $30$ dB attenuator, analog \ac{pa} output signals are collected by the \ac{vst} and then sent back to MATLAB.

The baseband signal $u(n)$ used for all experiments is an \ac{ofdm} signal with sampling frequency $200$ MHz, signal length $10^6$, and bandwidth $10$ MHz. An artificial \ac{iq} imbalance is added before sending the signal to the RF WebLab.  The gain imbalance is $1$ dB, and the phase imbalance is $\phi=8^{\circ}$. Frequency-dependent \ac{iq} imbalance is introduced using two $5$-th order \ac{fir} lowpass elliptic filters in the \ac{I} and \ac{Q} branches with different filter parameters: minimum stopband attenuation of $60$ dB (I) and $50$ dB (Q), the peak-to-peak ripples of $0.1$ dB (I) and $0.12$ dB (Q), the normalized passband edge frequencies of $0.8$ (I) and $0.85$ (Q). For more details of the frequency response difference between these two filters, refer to~\cite{zhu2013joint}.
The measured saturation point and measurement noise variance of the \ac{pa} in RF WebLab are $24.1$ V ($\approx 37.6$ dBm of a $50$ $\Omega$ impedance) and $0.0032$, respectively. The output signal of the \ac{pa} has an average power of $24.93$ dBm, which corresponds to a theoretical \ac{nmse} minimum~\cite{chani2018lower} of $-39.56$ dB and a simulated \ac{acpr} minimum~\cite{chani2018lower} of $-49.92$ dBc.
\subsubsection{Metrics}
We measure performance in terms of \ac{nmse} and \ac{acpr}. The \ac{nmse} is defined as 
\begin{align}
        \text{NMSE} = 10 \log_{10} \frac{ \mathbb{E}[|y(n) - u(n)|^2]}{\mathbb{E}[|u(n)|^2]},
    \label{eq:NMSE}
\end{align}
and gives the all-band error in time-domain between the \ac{pa} output signal and the \ac{dpd} input signal. The \ac{acpr} is defined as
\begin{align}
\displaystyle
        \text{ACPR} = 10 \log_{10}  \frac{\int_{\text{adj.}} |Y(f)|^2\text{d}f}{\int_{\text{ch.}} |Y(f)|^2\text{d}f},
    \label{eq:ACPR}
\end{align}
where $Y(f)$ denotes the Fourier transform of the \ac{pa} output signal. The integration in the numerator and denominator is performed over one adjacent channel (the one with larger integration between the lower and upper adjacent channel) and the main channel, respectively. The \ac{acpr} evaluates the amount of out-of-band emission.
\subsubsection{Benchmarks}
For a fair comparison, we consider the extended \ac{ph}~\cite{Lauri_PH_2010} because it is designed to jointly mitigate frequency-dependent \ac{iq} imbalance and \ac{pa} nonlinearity. Other referred Volterra-based models~\cite{ding2004MP, zhu2006dynamic, GMP_2006} fail to address both impairments. We also consider four other \ac{rvtdnn}-based models for comparison, namely \ac{rvtdnn}~\cite{liu2004dynamic}, \ac{rvftdnn}~\cite{Mee_2012}, \ac{arvtdnn}~\cite{wang2018augmented}, and \ac{r2tdnn}~\cite{wu2020residual}. All models use \ac{ila} for \ac{dpd} identification. All \ac{rvtdnn}-based models including the proposed \ac{arden} use the back-propagation algorithm with the Adam optimizer~\cite{kingma2014adam}, the \ac{mse} loss function, the ReLU activation function, and a mini-batch size of $256$. The extended \ac{ph}~\cite{Lauri_PH_2010} uses the least squares algorithm for parameter identification, and its computation complexity is given in Appendix~\ref{section:Appendix}.

\subsection{Results}\label{subsection:Results}
\subsubsection{Performance versus Complexity}
Fig.\,\ref{fig:nmse_flops} and Fig.\,\ref{fig:acpr_flops} show the \ac{nmse} and \ac{acpr} as a function of the number of \acp{flop} for the extended \ac{ph}, \ac{rvtdnn}, \ac{rvftdnn}, \ac{arvtdnn}, \ac{r2tdnn}~\cite{wu2020residual}, the proposed non-pruned~\ac{arden}, and \ac{arden} with a pruning factor $\eta_{\text{d}}=0.5$. For a fair comparison, all above \ac{dpd} schemes have memory length $3$, i.e., $M=3$ for \ac{arden}.  For the \ac{nn}-based structures, the number of \acp{flop} increases as the number of neurons in each hidden layers increases.  Specifically, we set the same number of hidden layers (three) for \ac{r2tdnn} and \ac{arden}, and the same number of neurons in each hidden layer, i.e., $K=5$ and $D_2=D_3=D_4$. Pruned \ac{arden} is based on the same structure of non-pruned \ac{arden}. The \ac{arvtdnn} contains three augmented envelope terms of the input signal (amplitude and its square and cube)~\cite[Tab.~II entry 11]{wang2018augmented} at the input layer. For \ac{ph}, the best results are selected with respect to the number of \acp{flop} through an exhaustive search of different values of its nonlinear order and filter length.

The proposed \ac{arden} with and without pruning achieves lower \ac{nmse} and \ac{acpr} results than all other \ac{dpd} schemes for all number of \acp{flop}. Specifically, the \ac{ph} has limited mitigation performance, flattens around a  \ac{nmse} of $-29.9$ dB and an \ac{acpr} of $-37.2$ dBc, whereas \ac{arden} achieves a \ac{nmse} of $-37.0$ dB and an \ac{acpr} of $-45.1$ dBc. Compared with the \ac{r2tdnn}~\cite{wu2020residual}, \ac{arden} yields sizable \ac{nmse} and \ac{acpr} gains for a number of \acp{flop} smaller than $500$, which verifies the effectiveness of the attention weights in the shortcut connections. Furthermore, \ac{arden} with a pruning factor $\eta_{\text{d}}=0.5$ requires even less number of \acp{flop} to achieve the same \ac{nmse} and \ac{acpr} compared with the non-pruned \ac{arden}, though this advantage vanishes as the size of \ac{arden} becomes large (\acp{flop}$>3000$).
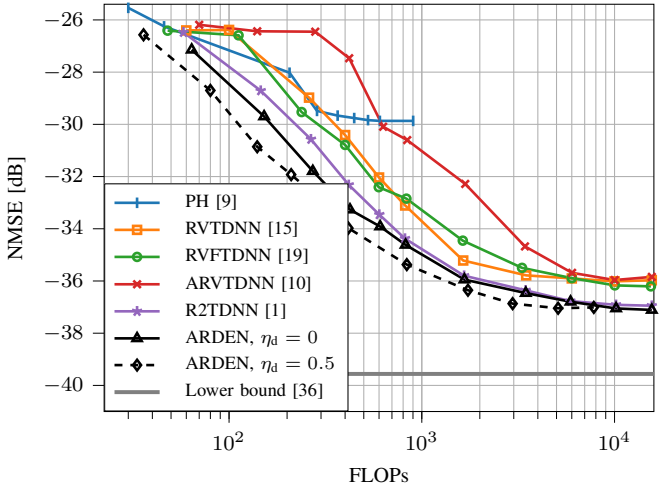
\begin{figure}[t]
    \centering
\begin{tikzpicture}[font=\footnotesize]

\newcommand\w{8.89}
\newcommand\h{7}
\definecolor{color0}{rgb}{0.12156862745098,0.466666666666667,0.705882352941177}
\definecolor{color1}{rgb}{1,0.498039215686275,0.0549019607843137}
\definecolor{color2}{rgb}{0.172549019607843,0.627450980392157,0.172549019607843}
\definecolor{color3}{rgb}{0.83921568627451,0.152941176470588,0.156862745098039}
\definecolor{color4}{rgb}{0.580392156862745,0.403921568627451,0.741176470588235}
\definecolor{color5}{rgb}{0.549019607843137,0.337254901960784,0.294117647058824}
\definecolor{color6}{rgb}{0.890196078431372,0.466666666666667,0.76078431372549}

\begin{axis}[
width=\w cm,
height=\h cm,
legend cell align={left},
legend style={at={(0,0)}, anchor=south west,fill opacity=1, draw opacity=1, text opacity=1},
log basis x={10},
tick align=outside,
tick pos=left,
x grid style={white!69.0196078431373!black},
xlabel={FLOPs},
xmajorgrids,
xmin=22.4846632250863, xmax=16000,
xminorgrids,
xmode=log,
xtick style={color=black},
y grid style={white!69.0196078431373!black},
ylabel={NMSE [dB]},
ymajorgrids,
ymin=-41, ymax=-25.4,
yminorgrids,
ytick style={color=black},
ytick distance={2}
]

\addplot [semithick, color0, line width=1.0pt, mark=|, mark size=2, mark options={solid}]
table {%
30 -25.5379108177306
46 -26.2464760888076
206 -28.0226642681877
286 -29.4853683167098
366 -29.6652441777166
446 -29.7595007130353
526 -29.8385611864743
606 -29.8674214430088
902 -29.8682348442571
};
\addlegendentry{\scriptsize \ac{ph}~\cite{Lauri_PH_2010}}
\addplot [semithick, color1, line width= 1.0pt, mark=square, mark size=1.5, mark options={solid}]
table {%
60 -26.4028737208903
100 -26.3835978495564
260 -28.9773659071555
400 -30.4104953415007
600 -32.0362199062596
820 -33.108022808425
1640 -35.2155999134006
3480 -35.7848570065807
6000 -35.9015141762483
10000 -36.0135792932070
15600 -35.9715102510826
};
\addlegendentry{\scriptsize \ac{rvtdnn}~\cite{liu2004dynamic}}
\addplot [semithick, color2, line width= 1.0pt, mark=o, mark size=1.5, mark options={solid}]
table {%
48 -26.4111290350099
112 -26.5902146823031
238 -29.5293396343662
400 -30.7878118017184
598 -32.4069726900535
832 -32.8458179594066
1632 -34.4544168280002
3312 -35.4969850735185
6000 -35.8908367424232
10032 -36.1682618174175
15438 -36.2059884761979
};
\addlegendentry{\scriptsize \ac{rvftdnn}~\cite{Mee_2012}}

\addplot [semithick, color3, line width=1.0pt, mark=x, mark size=2, mark options={solid}]
table {%
70 -26.1864170273475
140 -26.4350653599807
280 -26.4512767517426
420 -27.4675332706101
630 -30.0917188854929
840 -30.6031433025944
1680 -32.2825249681926
3430 -34.6758433727086
6020 -35.6879742367408
10010 -35.9638014502116
15610 -35.8427810193733
};
\addlegendentry{\scriptsize \ac{arvtdnn}~\cite{wang2018augmented}}

\addplot [semithick, color4,line width=1.0pt, mark=star, mark size=2, mark options={solid}]
table {%
58 -26.4740614843067
146 -28.7105497884385
266 -30.5710959062167
418 -32.3189183094985
602 -33.4573586109096
818 -34.3765026375705
1658 -35.7979007048369
3458 -36.3587001387286
5906 -36.7720871851811
10178 -36.9130001659492
15602 -36.9478469324459
};
\addlegendentry{\scriptsize R2TDNN~\cite{wu2020residual}}

\addplot [semithick, black, line width=1.0pt, mark=triangle, mark size=2, mark options={solid}]
table {%
64 -27.1433191154586
152 -29.6987576139656
272 -31.7930887921754
424 -33.2660906244235
608 -33.9144571479158
824 -34.6176467242217
1664 -35.9476837274451
3464 -36.4632855964073
5912 -36.7954570341793
10185 -37.0522597971072
15608 -37.1089711846626
};
\addlegendentry{\scriptsize \ac{arden}, $\eta_{\text{d}}=0$}
\addplot [semithick, black, dashed, line width=1.0pt, mark=diamond, mark size=2,, mark options={solid}]
table {%
36 -26.5702462920661
80 -28.7018569910725
140 -30.8583218100052
210 -31.9260673801021
308 -32.9870935771364
416 -33.9750166498039
836 -35.3720771733861
1736 -36.3535721708887
2960 -36.8676553914252
5096 -37.0444198304012
7808 -37.0118104924543
};
\addlegendentry{\scriptsize \ac{arden}, $\eta_{\text{d}}=0.5$}

\addplot [thick, color=gray, line width=1.5pt]
table {%
1 -39.56
16000 -39.56
};
\addlegendentry{\scriptsize Lower bound~\cite{chani2018lower}}
\end{axis}

\end{tikzpicture}
    \caption{\ac{nmse} as a function of the number of \acp{flop} for a DPD  of memory length $3$. The markers for \ac{ph}~\cite{Lauri_PH_2010} correspond to different sets of nonlinear order. The markers for \ac{rvtdnn}~\cite{liu2004dynamic},
    \ac{rvftdnn}~\cite{Mee_2012},  \ac{arvtdnn}~\cite{wang2018augmented}, and \ac{arden} correspond to different numbers of neurons in the hidden layers. For \ac{arden}, $K=5$ and $D_2=D_3=D_4$.}
    \label{fig:nmse_flops}
\end{figure}
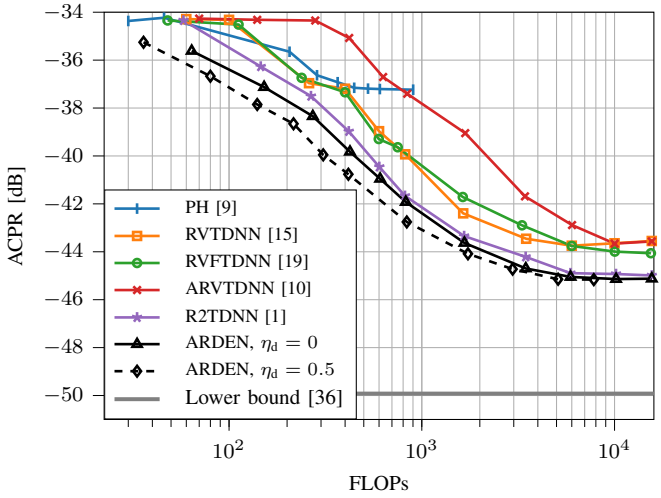
\begin{figure}[t]
    \centering
\begin{tikzpicture}[font=\footnotesize]
\newcommand\w{8.89}
\newcommand\h{7}
\definecolor{color0}{rgb}{0.12156862745098,0.466666666666667,0.705882352941177}
\definecolor{color1}{rgb}{1,0.498039215686275,0.0549019607843137}
\definecolor{color2}{rgb}{0.172549019607843,0.627450980392157,0.172549019607843}
\definecolor{color3}{rgb}{0.83921568627451,0.152941176470588,0.156862745098039}
\definecolor{color4}{rgb}{0.580392156862745,0.403921568627451,0.741176470588235}
\definecolor{color5}{rgb}{0.549019607843137,0.337254901960784,0.294117647058824}
\definecolor{color6}{rgb}{0.890196078431372,0.466666666666667,0.76078431372549}

\begin{axis}[
width=\w cm,
height=\h cm,
legend cell align={left},
legend style={at={(0,0)}, anchor=south west,fill opacity=1, draw opacity=1, text opacity=1},
log basis x={10},
tick align=outside,
tick pos=left,
x grid style={white!69.0196078431373!black},
xlabel={FLOPs},
xmajorgrids,
xmin=22.4846632250863, xmax=16000,
xminorgrids,
xmode=log,
xtick style={color=black},
y grid style={white!69.0196078431373!black},
ylabel={ACPR [dB]},
ymajorgrids,
ymin=-51, ymax=-34,
yminorgrids,
ytick style={color=black},
ytick distance={2}
]

\addplot [semithick, color0, line width=1.0pt, mark=|, mark size=2, mark options={solid}]
table {%
30 -34.3656774768395
46 -34.2310258030201
206 -35.6501915279622
286 -36.6311790555328
366 -36.9219390293242
446 -37.1548601651846
526 -37.1906551316222
606 -37.2072471370176
902 -37.2337595509424
};
\addlegendentry{\scriptsize \ac{ph}~\cite{Lauri_PH_2010}}
\addplot [semithick, color1, line width=1.0pt, mark=square, mark size=1.5, mark options={solid}]
table {%
60 -34.2878737499016
100 -34.3188525970525
260 -36.9627770450895
400 -37.1932452743455
600 -38.9665633288984
820 -39.9314681071262
1640 -42.3958726830312
3480 -43.4563751051804
6000 -43.7460371946858
10000 -43.64506449679877
15600 -43.54960208697903
};
\addlegendentry{\scriptsize \ac{rvtdnn}~\cite{liu2004dynamic}}
\addplot [semithick, color2, line width=1.0pt, mark=o, mark size=1.5, mark options={solid}]
table {%
48 -34.3446944509223
112 -34.5150683478444
238 -36.7389971993266
400 -37.3426977200394
598 -39.2932961198157
750 -39.636952056891
1632 -41.7135973124359
3312 -42.8953037534719
6000 -43.7577641245836
10030 -43.9924588569281
15438 -44.0604167299313
};
\addlegendentry{\scriptsize \ac{rvftdnn}~\cite{Mee_2012}}
\addplot [semithick, color3, line width=1.0pt, mark=x, mark size=2, mark options={solid}]
table {%
70 -34.2673742343647
140 -34.3171015557595
280 -34.3465529620829
420 -35.0754016124483
630 -36.7053963858387
840 -37.4023850364991
1680 -39.0476934751553
3430 -41.6861199481789
6020 -42.8845044335789
10010 -43.6583989860362
15610 -43.5674860217834
};
\addlegendentry{\scriptsize \ac{arvtdnn}~\cite{wang2018augmented}}

\addplot [semithick, color4, line width=1.0pt, mark=star, mark size=2, mark options={solid}]
table {%
58 -34.3438404645452
146 -36.2692778064091
266 -37.5037174269115
418 -38.9727693753292
602 -40.4579453150425
818 -41.6658084826266
1658 -43.3436041570119
3458 -44.2135631352436
5906 -44.8925945610758
10178 -44.9287287652953
15602 -44.988640570607224
};
\addlegendentry{\scriptsize R2TDNN~\cite{wu2020residual}}

\addplot [semithick, black, line width=1.0pt, mark=triangle, mark size=2, mark options={solid}]
table {
64 -35.6154980335459
152 -37.1259215447865
272 -38.3416821164422
424 -39.8259803604112
608 -40.9554993759642
826 -41.9228959745517
1664 -43.6309202010196
3464 -44.6967296301151
5912 -45.0500031851986
10184 -45.1385274146045
15608 -45.1234849509054
};
\addlegendentry{\scriptsize \ac{arden}, $\eta_{\text{d}}=0$}
\addplot [semithick, black, dashed, line width=1.0pt, mark=diamond, mark size=2, mark options={solid}]
table {%
36 -35.2540909506771
80 -36.6719097634597
140 -37.8509737942403
216 -38.6566641256403
308 -39.9537915986086
416 -40.7642871587075
836 -42.7547671724047
1736 -44.0842271793729
2960 -44.7238402520233
5096 -45.1459518177599
7808 -45.1622606321834
};
\addlegendentry{\scriptsize \ac{arden}, $\eta_{\text{d}}=0.5$}

\addplot [thick, color=gray, line width=1.5pt]
table {%
1 -49.92
16000 -49.92
};
\addlegendentry{Lower bound~\cite{chani2018lower}}
\end{axis}
\end{tikzpicture}
    \caption{\ac{acpr} as a function of the number of \acp{flop} for a DPD of memory length $3$. The markers for \ac{ph}~\cite{Lauri_PH_2010} correspond to different sets of nonlinear order. The markers for \ac{rvtdnn}~\cite{liu2004dynamic},
    \ac{rvftdnn}~\cite{Mee_2012},  \ac{arvtdnn}~\cite{wang2018augmented}, and \ac{arden} correspond to different numbers of neurons in the hidden layers. For \ac{arden}, $K=5$ and $D_2=D_3=D_4$.}
    \label{fig:acpr_flops}
\end{figure}

\subsubsection{Complexity-Restricted Scenario}
\begin{figure}[t]
    \centering
    \input{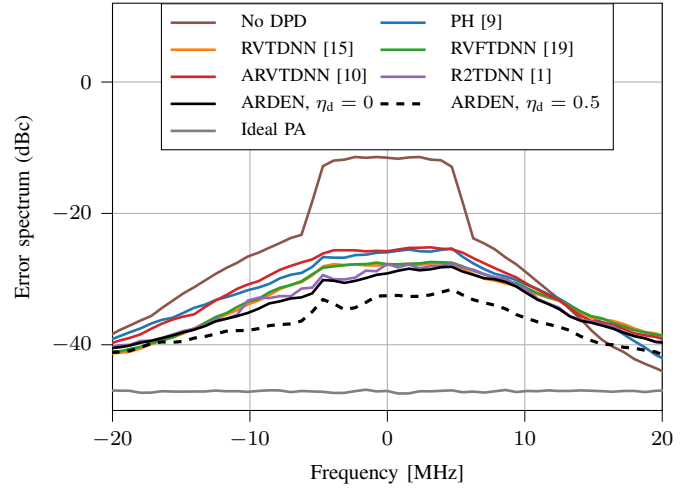}
    \label{fig:psd_err_w_iqi}
    \caption{Error spectrum between the actual and desired \ac{pa} output signals for different models in a computation-restricted scenario with around $400$ \acp{flop}.}
    \label{fig:psd_err_iqi}
\end{figure}
\begin{table}[t]
\centering
\caption{\ac{nmse} and \ac{acpr} results of  the \ac{ph}~\cite{Lauri_PH_2010}, \ac{rvftdnn}~\cite{Mee_2012} \ac{rvtdnn}~\cite{liu2004dynamic}, \ac{arvtdnn}~\cite{wang2018augmented}, and proposed~\ac{arden} in Fig.\,\ref{fig:psd_err_iqi}. The lower bound results are the minimum that can be achieved at an average output power $25.19$ dBm.}
\begin{tabular}{@{}cccc@{}}
\toprule
        & \acp{flop} & \ac{nmse} [dB] &  \ac{acpr} [dBc] \\ \midrule
No \ac{dpd} &---  & $-17.84$    & $-34.40$  \\ 

\ac{ph}~\cite{Lauri_PH_2010}      &$446$                                                            &$-29.69$     &$-36.76$                                                               \\
\ac{rvtdnn}~\cite{liu2004dynamic}  &$400$                                                           &$-31.13$        &$-37.91$                                                                    \\
\ac{rvftdnn}~\cite{Mee_2012} &$400$                                                                        &$-30.93$      &$-37.54$                                                                 \\
\ac{arvtdnn}~\cite{wang2018augmented} & $420$                                                              &$-29.22$        & $-36.39$                                                                  \\
\ac{r2tdnn}~\cite{wu2020residual} & $418$                                                              &$-32.06$        & $-38.01$                                                                \\
\ac{arden}, $\eta_{\text{d}}=0$  &$424$       &$-33.26$                                                                            & $-38.97$                        \\ 
\ac{arden}, $\eta_{\text{d}}=0.5$  &$416$   &$\mathbf{-34.58}$  &$\mathbf{-41.82}$ \\
Lower bound~\cite{chani2018lower} & ---   &$-39.56$    &$-49.92$  \\ 
\bottomrule
\end{tabular}
\label{tab:psd_results}
\end{table}
We compare the mitigation performance of different \ac{dpd} schemes in a limited complexity scenario for a number of \acp{flop} around $400$. Fig.\,\ref{fig:psd_err_iqi} shows the error spectrum of the \ac{pa} output without \ac{dpd}, with \ac{dpd} via \ac{ph}, \ac{rvtdnn}, \ac{rvftdnn}, \ac{arvtdnn}, \ac{arden}, pruned \ac{arden}, and of an ideal linear \ac{pa}. The corresponding number of \acp{flop}, \ac{nmse}, and \ac{acpr} results are given in Table \ref{tab:psd_results}. The pruned \ac{arden} is based on an original \ac{arden} with $C_{\text{arden}}=818$ \acp{flop} and a pruning factor $\eta_{\text{d}}=0.5$. For a fair comparison, the memory length for all \ac{dpd} schemes is set to $3$, and the number of \acp{flop} for each scheme is $\approx 400$ by adjusting the number of neurons in the hidden layers for \ac{nn}-based schemes and the nonlinear order for \ac{ph}. As shown in Fig\,\ref{fig:psd_err_iqi}, without \ac{dpd}, there are considerable in-band and out-of-band distortions, which are not fully compensated by any of the \ac{dpd} schemes due to residual unrecoverable distortions in the \ac{iq}-\ac{pa} system. The pruned \ac{arden} with $\eta_{\text{d}}=0.5$ achieves the best performance with \ac{nmse} of $-34.58$ dB and \ac{acpr} of $-41.82$ dB, while requiring a similar number of \acp{flop}.

\subsubsection{Interpretation of Pruning and Attention}\label{section:result_prune}
\begin{figure}[t]
    \centering
    \input{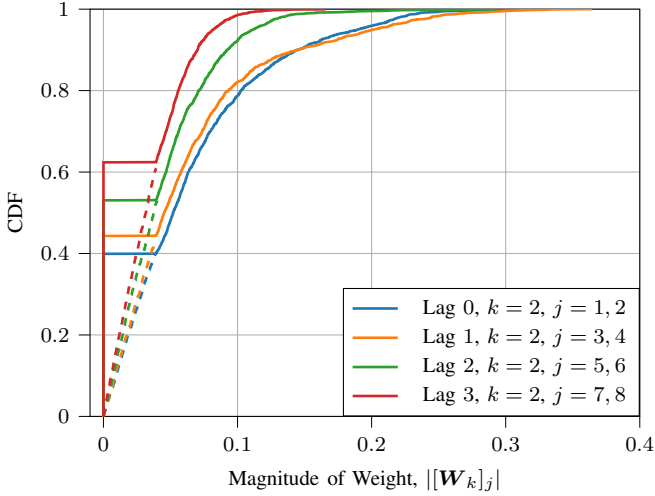}
    \caption{CDFs of the weight magnitudes for neurons in the input layer of \ac{arden} fed with input signals with lag $0$, $1$, $2$, and $3$ before and after pruning (dashed and solid lines), respectively. $K=3$, $D_2=512$, and $\eta_{\text{d}}=0.5$. The final remaining weights have a minimal magnitude around $0.04$.}
    \label{fig:nmse_prune_cdf}
\end{figure}
Considering \ac{arden} with $K=3$, $M=3$, $D_2=512$, and $\eta_{\text{d}}=0.5$,  Fig.\,\ref{fig:nmse_prune_cdf} shows the \acp{cdf} of the magnitude of weights in the connections between the first and second layers, i.e., $|\boldsymbol{W}_2|$, before and after pruning. Each \ac{cdf} corresponds to the the magnitude of weights in the connections for every two neurons in the first layer fed with input signals of lag $0$, $1$, $2$, and $3$, i.e., $|[\boldsymbol{W}_2]_{1,2}|$, $|[\boldsymbol{W}_2]_{3,4}|$, $|[\boldsymbol{W}_2]_{5,6}|$, and $|[\boldsymbol{W}_2]_{7,8}|$, respectively. The dashed and solid lines correspond to before and after pruning, respectively. The remaining weights have a minimal magnitude around $0.04$.

Note that the weights in the connections for neurons fed with shorter lag input signals have larger magnitudes ($>0.04$) than for neurons fed with longer lag input signals, especially for lag $0$. Thus, despite the pruning factor $\eta_{\text{d}}=0.5$ for the second layer, more weights ($>50\%$) are masked to zero for neurons with longer lags than for neurons of shorter lags ($<50\%$). This indicates the presence of an inherent attention mechanism during the impairment mitigation of the \ac{iq}-\ac{pa} system.

\subsubsection{Large Sparse versus Small Dense}\label{section:results_sparse_vs_dense}
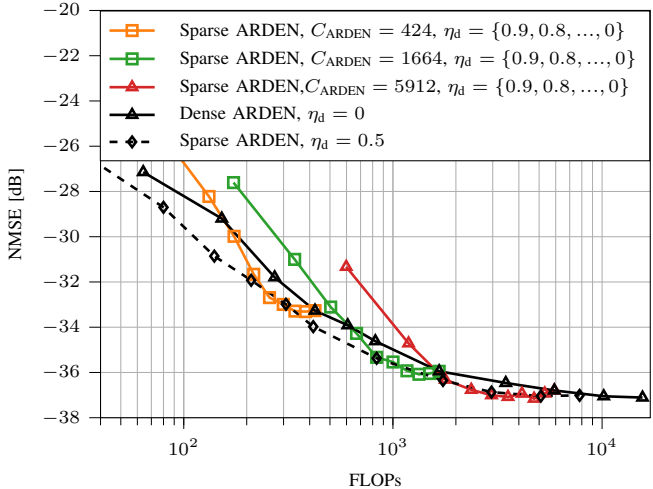
\begin{figure}[t]
    \centering
\begin{tikzpicture}[font=\scriptsize]
\newcommand\w{8.89}
\newcommand\h{7}
\definecolor{color0}{rgb}{0.12156862745098,0.466666666666667,0.705882352941177}
\definecolor{color1}{rgb}{1,0.498039215686275,0.0549019607843137}
\definecolor{color2}{rgb}{0.172549019607843,0.627450980392157,0.172549019607843}
\definecolor{color3}{rgb}{0.83921568627451,0.152941176470588,0.156862745098039}
\definecolor{color4}{rgb}{0.580392156862745,0.403921568627451,0.741176470588235}
\definecolor{color5}{rgb}{0.549019607843137,0.337254901960784,0.294117647058824}
\definecolor{color6}{rgb}{0.890196078431372,0.466666666666667,0.76078431372549}

\begin{axis}[
width=\w cm,
height=\h cm,
legend cell align={left},
legend style={at={(1,1)}, anchor=north east,fill opacity=1, draw opacity=1, text opacity=1},
tick align=outside,
tick pos=left,
x grid style={white!69.0196078431373!black},
xlabel={FLOPs},
xmode=log,xmajorgrids,xminorgrids,
xmin=40, xmax=17000,
xtick style={color=black},
y grid style={white!69.0196078431373!black},
ylabel={NMSE [dB]},
ymin=-38, ymax=-20,
ytick style={color=black},
ymajorgrids,
yminorgrids,
ytick style={color=black},
ytick distance={2}
]

\addplot [semithick, color1, mark=square, line width=1.0pt, mark size=2, mark options={solid}]
table {%
   49  -23.7021
   91  -26.2322
  132  -28.2218
  174  -29.9825
  216  -31.6596
  257  -32.6907
  299  -32.9921
  340  -33.2885
  382  -33.3109
  424  -33.2661
};
\addlegendentry{Sparse ARDEN, $C_{\text{ARDEN}}=424$, $\eta_{\text{d}}=\{0.9,0.8,...,0\}$}

\addplot [semithick, color2, mark=square, line width=1.0pt, mark size=2, mark options={solid}]
table {%
173	-27.6076694711118
339	-31.0020811420646
504	-33.1090359149173
670	-34.2737192407713
836	-35.3307993932062
1001	-35.5381583660613
1167	-35.9259605029726
1332	-36.0805191177696
1498	-36.0441625138512
1664	-35.9476837274451
};
\addlegendentry{Sparse ARDEN, $C_{\text{ARDEN}}=1664$, $\eta_{\text{d}}=\{0.9,0.8,...,0\}$}
\addplot [semithick, color3, mark=triangle, line width=1.0pt, mark size=2.2, mark options={solid}]
table {%
598	-31.3301017766603
1188	-34.7127730150297
1779	-36.3172562398456
2369	-36.7657373643601
2960	-37.0051326430188
3550	-37.0789232941872
4140	-36.9347166557972
4731	-37.1580276660343
5321	-36.9267185609696
5912	-36.7954570341793
};
\addlegendentry{Sparse ARDEN,$C_{\text{ARDEN}}=5912$, $\eta_{\text{d}}=\{0.9,0.8,...,0\}$}

\addplot [semithick, black, line width=1.0pt, mark size=2.2, mark=triangle,  mark options={solid}]
table {%
64 -27.1433191154586
152 -29.1987576139656
272 -31.7930887921754
424 -33.2660906244235
608 -33.9144571479158
824 -34.6176467242217
1664 -35.9476837274451
3464 -36.4632855964073
5912 -36.7954570341793
10185 -37.0522597971072
15608 -37.1089711846626
};
\addlegendentry{Dense ARDEN,  $\eta_{\text{d}}=0$}

\addplot [semithick, black, dashed, line width=1.0pt, mark=diamond, mark size=2,, mark options={solid}]
table {%
36 -26.5702462920661
80 -28.7018569910725
140 -30.8583218100052
210 -31.9260673801021
308 -32.9870935771364
416 -33.9750166498039
836 -35.3720771733861
1736 -36.3535721708887
2960 -36.8676553914252
5096 -37.0444198304012
7808 -37.0118104924543
};
\addlegendentry{Sparse \scriptsize ARDEN, $\eta_{\text{d}}=0.5$}

\end{axis}

\end{tikzpicture}
    \caption{NMSE as a function of the number of \acp{flop} for dense and sparse \acp{arden} with fixed $\eta_{\text{d}}=\{0,0.5\}$ and varied $\eta_{\text{d}}=\{0.9,0.8,...,0\}$. The markers for dense and sparse \acp{arden} correspond to different number of neurons in the hidden layers as in Fig.~\ref{fig:nmse_flops} and  different $\eta_{\text{d}}$, respectively. }
    \label{fig:nmse_prune_spar}
\end{figure}
The performance of sparse and dense \acp{arden} is compared in Fig.\,\ref{fig:nmse_prune_spar}. It illustrates the \ac{nmse} as a function of the number of \acp{flop} for three sparse \acp{arden} with a varied $\eta_{\text{d}}=\{0.9,0.8,...,0\}$ but different non-pruned complexity $C_{\text{ARDEN}}=\{424, 1664, 5912\}$, and two \acp{arden} with fixed $\eta_{\text{d}}=\{0, 0.5\}$ from Fig.~\ref{fig:nmse_flops}. All the \acp{arden} have the same number of layers $K=5$.

For a given number of \acp{flop}, sparse \acp{arden} allow to outperform the dense \acp{arden}. A larger size \ac{arden} ($C_{\text{ARDEN}}=5912$) allows a larger $\eta_{\text{d}}=0.7$ than that of a smaller size \ac{arden} ($C_{\text{ARDEN}}=424$) with $\eta_{\text{d}}=0.5$. This indicates that there is an optimal pruning factor for a given sized \ac{arden}. The \ac{arden} with a fixed $\eta_{\text{d}}=0.5$ performs nearly always the best over all complexities, which suggests that it is better to train a $2\times$ larger size dense \ac{arden} and prune it to the desired complexity than using the best dense \ac{arden}.

\section{Conclusion}\label{section:conclusion}
We proposed a novel attention residual learning \ac{nn}, referred to as \ac{arden}, for low-complexity mitigation of multiple hardware impairments in direct conversion transmitters. \ac{arden} keeps the instantaneous linear input-output relation of the transmitter by adding two trainable neuron-wise shortcut connections between the corresponding neurons of the input and output layers. Furthermore, we proposed and analyzed a \ac{nn} connection pruning algorithm, which allows to gradually remove weights of minimum magnitude in each layer. Experimental results show that \ac{arden} with a pruning factor of $0.5$ achieves a \ac{nmse} gain $>2.5$ dB and an \ac{acpr} gain $>2$ dBc compared to other \ac{rvtdnn}-based models and a Volterra-based model proposed in the literature, with less or similar complexity.

\appendices
\section{Computation Complexity of the PH}\label{section:Appendix}
The extended PH~\cite{Lauri_PH_2010} is based on the PH model~\cite{ding2004MP} given by the polynomials
\begin{equation}
    \psi_p(x(n)) = \sum_{k\in I_p}a_{k,p}|x(n)|^{k-1} x(n), p\in  I_p,
    \label{eq:PH_poly}
\end{equation}
where $p$ is the polynomial order, $I_p=\{1,3,...,p\}$ for only odd orders, and $a_{k,p}$ are the polynomial weights.  The polynomial~\eqref{eq:PH_poly} and its conjugate $\psi_p(x^{\ast}(n))$ are filtered by \ac{fir} filters $h_{p}(n)$ and $h_{q}(n)$ of length $L_p$ and $L_q$, respectively. The output of the extended PH is 
\begin{equation}
    y(n) = \sum_{p=1}^{P}h_{p}(n)\circledast \psi_p(x^*(n)) + \sum_{q=1}^{Q}h_{q}(n)\circledast\psi_q(x^*(n)),
    \label{eq:PH_overall}
\end{equation}
where $P$ and $Q$ are the polynomial orders for the non-conjugate and conjugate branches and $\circledast$ denotes convolution. 

The number of complex-valued weights in~\eqref{eq:PH_poly} is  
\begin{equation}
    N_{\text{PH, poly}} = \left(1+\frac{P+1}{2}\right)\frac{P+1}{4} + \left(1+\frac{Q+1}{2}\right)\frac{Q+1}{4}.
    \label{eq:PH_num_pw}
\end{equation}
$8$ \acp{flop} are required for each weight, where $6$ \acp{flop} are for the complex multiplication and $2$ \acp{flop} for the complex summation~\cite{Complexity_Ali_2010}. Similarly, the number of complex-valued filter parameters is~\cite{Lauri_PH_2010} 
\begin{equation}
  N_{\text{PH, filter}} = \sum_{p\in I_P}L_p + \sum_{q\in I_Q}L_q + 1,
  \label{eq:PH_num_filter}
\end{equation}
which also require $8$ \acp{flop} each. In total, the number of \acp{flop} required for the \ac{ph} is
\begin{equation}
\begin{aligned}
        C_{\text{PH}} &= 8(N_{\text{PH, poly}} + N_{\text{PH, filter}})-4 + 3+(\max(P,Q)-1). 
     \end{aligned}
\end{equation}
 \balance

\bibliographystyle{IEEEtran}
\bibliography{main}

\end{document}